\documentclass[iop]{emulateapj}
\usepackage{amsmath}
\usepackage{commath}

\begin{document}


\title{Dust coagulation in the vicinity of a gap-opening Jupiter-mass planet }


\author{Augusto Carballido\altaffilmark{1}, Lorin S. Matthews\altaffilmark{1}, and Truell W. Hyde\altaffilmark{1}}
\affil{Center for Astrophysics, Space Physics and Engineering Research, Baylor University, Waco, TX 76798, USA}
\email{Augusto\_Carballido@baylor.edu}






\begin{abstract}
 We analyze the coagulation of dust in and around a gap opened by a Jupiter-mass planet. To this end, we carry out a high-resolution magnetohydrodynamic (MHD) simulation of the gap environment, which is turbulent due to the magnetorotational instability. From the MHD simulation, we obtain values of the gas velocities, densities and turbulent stresses a) close to the gap edge, b) in one of the two gas streams that accrete onto the planet, c) inside the low-density gap, and d) outside the gap. The MHD values are then supplied to a Monte Carlo dust coagulation algorithm, which models grain sticking and compaction. We consider two dust populations for each region: one whose initial size distribution is monodisperse, with monomer radius equal to 1 $\mu$m, and another one whose initial size distribution follows the Mathis-Rumpl-Nordsieck distribution for interstellar dust grains, with an initial range of monomer radii between 0.5 and 10 $\mu$m. Our Monte Carlo calculations show initial growth of dust aggregates followed by compaction in all cases but one, that of aggregates belonging to the initially monodisperse population subject to gas conditions outside the gap. In this latter case, the mass-weighted (MW) average porosity of the population reaches extremely high final values of 98\%. The final MW porosities in all other cases range between 30\% and 82\%. The efficiency of compaction is due to high turbulent relative speeds between dust particles. Future studies will need to explore the effect of different planet masses and electric charge on grains.
\end{abstract}





\section{Introduction}

Numerical investigations of the dynamics of dust particles in the neighborhood of a gap-opening planet reveal that they tend to concentrate at gap edges, for planet masses between $\sim 0.03 M_{\rm{J}}$ and $5 M_{\rm{J}}$, with $M_{\rm{J}}$ the mass of Jupiter \citep{fou07,zhuetal14,owen14,glez15,pk15}. These accumulations are of great interest not only because they could facilitate planetesimal formation, but also because they could supply solid material to the planetary atmosphere through deposition of small grains, and hence contribute to the atmospheric opacity, although perhaps modestly \citep{ormel14}. In particular, a Jupiter-mass planet accretes predominantly particles smaller than 10 $\mu$m \citep{paar07}, which places a tight constraint on its solid enrichment. 
 
One aspect of the evolution of dust in the vicinity of a gap-opening planet that has not been considered is the internal structure of solid aggregates. The porosity of dust particles affects their aerodynamic coupling to the disk gas. Their structure can also affect the ionization state of the protoplanetary disk: the increased surface area associated with internal voids within the dust aggregates allows for fast electron recombination, which can lower the ionization levels of the gas phase, and thus render the disk gas stable to the magnetorotational instability, or MRI \citep{bh91}, a likely candidate to produce turbulent viscosity in disks [although other mechanisms have been proposed to transfer a protoplanetary disk's angular momentum, such as centrifugally-driven winds and jets \citep{bp82} and magnetic breaking \citep{mt04}].

Dust in the vicinity of a gap-opening planet is subject to a dynamic and energetic environment. For example, inside the gap, shocks are generated when streams of gas flow through the L1 and L2 points towards the interior of the planet's Hill sphere \citep{lsa99}. These shocks can modify the crystalline structure of silicates through thermal annealing. Knowledge of the porous structure of dust aggregates is essential to determine their heat conductivity during these events.

Previous studies of dust coagulation and dust charging in protoplanetary disks show that electric charges lead to larger, more massive and more porous aggregates than in the case of neutral coagulation \citep{mlh12}. In turn, the evolution of the aggregate porosity may define a region in weakly ionized disks where the growth of sub-micron-sized particles becomes stalled due to electrostatic repulsion between negatively-charged aggregates \citep{otts11}.

The dusty ingredient in the MRI recipe could also be crucial for the dynamics of giant-planet circumplanetary disks. \citet{tls14} showed that, under some circumstances, the circumjovian disk has sufficient electrical conductivity for magnetic forces to drive accretion stresses. But if the disk contains enough dust, MRI turbulence is unlikely to occur, even in the presence of ionizing X-ray radiation. The circumjovian disk can develop a substantial magnetically inactive ``dead zone", where regular satellites could form.


It is thus evident that the size and internal structure of dust grains are key to understanding the birth environment of planets that are capable of opening gaps. To better characterize the spatial distribution of these dust properties around one such planet, in this study we combine a numerical model of MRI turbulence with a dust-coagulation algorithm. Our results will provide a clearer picture of the dust aggregate structure around a young, Jupiter-mass planet. In the interest of specificity, we use our models to invoke a scenario in which Jupiter underwent migration within the primitive solar nebula, the so-called Grand Tack hypothesis \citep{wmrom11}. According to this hypothesis, Jupiter migrated from inside its current position (but beyond the Main Asteroid Belt) to a distance of about 1.5 AU from the Sun, where it encountered an orbital resonance with the trailing Saturn. Both planets then reversed their orbital motion outwards. The Grand Tack helps reproduce the formation of Mars analogs with the correct mass from a disk of planetary embryos.

In Section \ref{sec:method} we describe our numerical models. Section \ref{sec:results} presents our results. In Section \ref{sec:discussion} we discuss these results, and we make concluding remarks in Section \ref{sec:conclusions}.

\section{Method}\label{sec:method}
\subsection{Disk model}
We use a magnetohydrodynamic (MHD) protoplanetary disk model in the local shearing box approximation \citep{hgb95}, in which the MHD equations are solved in a rectangular coordinate system that corotates with the disk at a fiducial orbital radius $R_0$, with angular frequency $\Omega(R_0)$. In this system, the $x$ axis is oriented along the radial direction, the $y$ axis along the azimuthal direction, and the $z$ axis is parallel to the disk's angular momentum vector. Our solver is the Athena code \citep{sgths08}, a grid-based algorithm that has been extensively tested and employed for various studies of protoplanetary disks. 

Our numerical setup is similar to that of \citet{zsr13}: the local, 3D disk model is isothermal, and does not include vertical stratification of the gas density. The box dimensions are $16H\times 16H\times H$, where $H$ is the disk scale height. We use a numerical resolution twice as large as that in \citet{zsr13}, namely 64 grid cells per $H$. In the code's system of units, the gas sound speed $c_{\rm s}$, the initial gas density $\rho_0$, and the angular frequency $\Omega$ are all equal to 1. The initial magnetic field strength is given by the plasma beta, $\beta=400$, and the initial field configuration corresponds to a non-zero net vertical magnetic flux. 

To model the gravitational effect of a planet on the surrounding disk gas, we place a cylindrical potential at the center of the box, with axis coincident with the box's vertical axis. As in \citet{zsr13}, the planet potential is given by

\begin{equation}
\Phi_{\rm p}=-GM_{\rm p}\frac{r^2 + 1.5r^2_{\rm s}}{\left(r^2 + r^2_{\rm s}\right)^{3/2}},
\end{equation}

\noindent where $G$ is the gravitational constant, $M_{\rm p}$ is the planet mass, $r$ is the distance to the $z$ axis, and $r_{\rm s}$ is a smoothing length to avoid small time steps close to the source. The potential is also smoothed out 
to take into account discontinuities at the box boundaries:

\begin{equation}
\Phi_{\rm p,s}\left(r\right) = \begin{cases} \Phi_{\rm p}\left(r_f\right)-\sqrt{\left[\Phi_{\rm p}\left(r_f\right)-\Phi_{\rm p}\left(r\right)\right]^{2} + G^{2}M^{2}_{\rm p}/r^{2}_{\rm co}}, \\
\hfill \text{if $r<r_f$}\\
\Phi_{\rm p}\left(r_f\right)-GM_{\rm p}/r_{\rm co}, \hfill \text{if $r > r_f$}
\end{cases}
\end{equation}

\noindent where $r_f$ is the cutoff distance to the $z$ axis beyond which the potential flattens out, and $r_{\rm co}$ is a smoothing length at the cutoff radius. Following \citet{zsr13}, we set $r_f=7.5H$ and $r_{\rm co}=50H$.

In Athena, we express the planet mass in terms of the so-called thermal mass $M_{\rm th}$, the mass at which the Hill radius and the Bondi radius of the planet are comparable to the disk scale height \citep{raf06}. For the minimum-mass solar nebula model [MMSN; \citet{hayashi81}], the thermal mass is \citep{drs11}

\begin{equation}\label{eq:mth}
M_{\rm th}\approx 12\left(\frac{c_{\rm s}}{1\, \rm{km\, s^{-1}}}\right)^{3}\left(\frac{R_{\rm p}}{1\, \rm{AU}}\right)^{3/4} M_{\earth},
\end{equation}

\noindent where $R_{\rm p}$ is the planet's semi-major axis, and $M_{\earth}$ is the mass of Earth. Within the Grand Tack scenario, Jupiter begins its inward migration at $\sim3.5$ AU, reaches 1.5 AU, and reverses its motion towards its current orbital position at 5.2 AU \citep{wmrom11}. Since we are not modeling the actual migration, but effectively only a ``snapshot" of that process, we choose $R_{\rm p}=3$ AU, because it is an interesting region of the solar nebula in terms of the dynamical evolution of chondritic parent bodies \citep{wmrom11}.  In that case, from relation (\ref{eq:mth}) a value of $6.64M_{\rm th}$ gives our planet a mass $M_{\rm p}\approx 1M_{\rm J}$. 

We run this high-resolution setup for 51 orbits (265 yr at 3 AU in the MMSN), and use the resulting data from the last 12 orbits, after the gas flow reaches an approximate steady state, as input to the coagulation code described next.

\subsection{Dust coagulation model}
Our treatment of dust coagulation follows the implementation of \citet{ost07}, which in turn is based on the Monte Carlo model by \citet{gill75}. A population of $N$ particles (we refer to monomers and aggregates jointly as particles) is assumed to be uniformly distributed inside an abstract volume $V$. The collision rate between particles $i$ and $j$ is calculated as 

\begin{equation}\label{eq:cij}
C_{ij}=\sigma_{ij}\Delta v_{ij}/V,
\end{equation}

\noindent where $\sigma_{ij} =\pi\left(a_i + a_j\right)^2$ is the collision cross-section, $a_i$ and $a_j$ are the respective particle radii, and $\Delta v_{ij}$ is the relative speed between the two particles (described below). In this coagulation algorithm, we track the radius of any particle $k$, $a_k$, through the enlargement factor $\psi_{k} = \mathcal{V}_k/\mathcal{V}_0$, where $\mathcal{V}_k$ is the extended volume of the particle (i.e., the volume corresponding to the geometric cross-section of the aggregate), and $\mathcal{V}_0$ is the volume of a monomer of radius $a_0$. We then have $a_k=a_0\psi_k^{1/3}$. We also track the evolution of the particle mass $m_k$.






The particle pair $(i,j)$ that will be involved in each collision is chosen using random numbers and partial sums of the collision rate $C_{ij}$. We treat two possible outcomes of a binary collision: particles either merely stick without further restructuring, or are compacted at the expense of internal voids. These possibilities occur when the kinetic energy of collision ($E$) is less than the critical energy to initiate compaction ($E_{\rm{comp}}$), or exceeds it, respectively \citep{ost07}. To conserve the total number $N$ of particles in the volume $V$ after a collision, one of the remaining $N-1$ particles is randomly chosen and duplicated. The duplication procedure requires that the volume $V$ be increased to preserve the spatial density of dust, $\rho_{\rm d}=\sum_{i}m_{i}/V$.

In order to include the effect of the evolving porosity of aggregates formed in collisions, the enlargement factor $\psi_k$ of an aggregate is calculated according to Eqs. (15) or (17) of \citet{ost07}, depending on whether $E<E_{\rm{comp}}$ or $E\geq E_{\rm{comp}}$, respectively.

\begin{figure}
\figurenum{1}
\epsscale{1.2}
\plotone{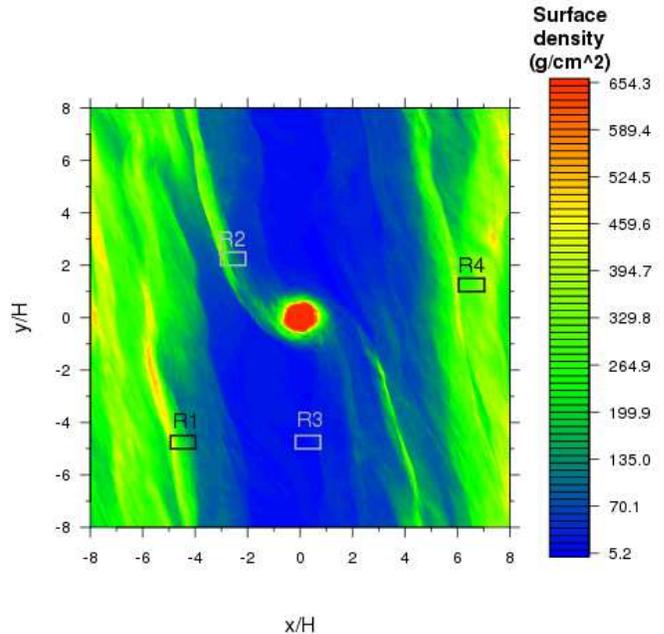}
\caption{Contours of gas surface density at the end of our MHD simulation ($t=51$ orbits, or 265 yr at 3 AU in the MMSN), obtained by averaging the gas density over the vertical extent of the shearing box. The rectangles labeled R1 through R4 mark the box locations from which gas velocities, densities and stresses are fed to the Monte Carlo coagulation code. }
\label{fig:xymap}
\end{figure}

The relative speed $\Delta v_{ij}$ in Eq. (\ref{eq:cij}) has contributions from Brownian motion and from turbulence. We use an expression for the turbulent relative speed based on Eq. (10) of \citet{oct08}. We insert into that expression values of the gas turbulent velocity, turbulent viscosity (given by magnetic and hydrodynamic stresses) and gas density, all obtained from the MHD simulation described above. These values are taken from four different regions in the vicinity of the planet, as shown in Fig. \ref{fig:xymap}, which portrays the system at the end of our MHD simulation. The contours represent the gas surface density, obtained by averaging the gas density over the vertical extent of the box. The gap opened by the planet is clearly defined. The regions R1 through R4 are each divided into 128 subregions of $4\times4$ grid cells each. We call these subregions macrocells. The gas variables are averaged inside each macrocell, and are then fed to the Monte Carlo coagulation code. Note that although the volume $V$ where each particle population resides is effectively \textit{associated} to each macrocell, it is \textit{not} the volume of the macrocell. The particles do not have position coordinates associated with them. 

We set the number of particles $N$ in the volume $V$ to $10^4$, and we use two different initial conditions for the particle sizes: $a_0$=1 $\mu$m, and $a_0$ distributed according to the Mathis-Rumpl-Nordsieck (MRN) distribution of dust grains in the interstellar medium, $n(a_0)\propto a_0^{-3.5}$ \citep{mrn}. In the latter case, the range of monomer radii is 0.5 $\mu \textrm{m} < a_0 < 10  \,\mu$m. In both cases, the monomer bulk density is $\rho_{\rm b}$=3 g cm$^{-3}$. 

\section{Results}\label{sec:results}
\begin{figure}
\figurenum{2}
\epsscale{1.2}
\plotone{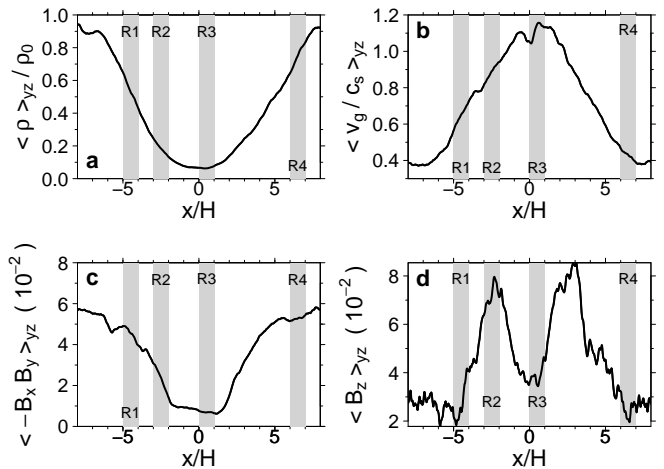}
\caption{Radial profiles of \textbf{(a)} gas density normalized by its initial value, \textbf{(b)} gas turbulent velocity in units of the sound speed $c_{\rm s}$, \textbf{(c)} $x$--$y$ component of magnetic stress tensor, and \textbf{(d)} vertical component of magnetic field. All profiles are time, $y$ and $z$ averages. Vertical gray bars denote the radial position of the reference regions in Fig. \ref{fig:xymap}}
\label{fig:radprofs}
\end{figure}

Figure \ref{fig:radprofs} shows radial profiles of various quantities associated with the MHD flow. These quantities have been averaged over time (the last 12 orbits of the simulation) and over the $y$ and $z$ directions (regions for which $\abs{y}<1$ are not included). The vertical gray bars mark the radial position of the reference regions. In Fig. \ref{fig:radprofs}a, the gas density, normalized by the initial density, exhibits a drop of a factor of $\sim10$ inside the gap opened by the planet with respect to the surrounding gas. Panel c shows a corresponding factor-of-8 drop in the Maxwell stress, which is the main contributor to the turbulent viscosity in accretion disks \citep{hgb95}. Figure \ref{fig:radprofs}b shows the turbulent gas velocity in units of the gas sound speed. Overall, turbulent gas speeds are relatively high, reflecting the turbulent strength for our case of a non-vanishing initial magnetic flux. Finally, Fig. 2d reveals that the mean vertical magnetic field has two maxima at either side of the planet's radial position, close to the gap edges. This is in contrast to the centrally-peaked profile of $\langle B_z\rangle_{yz}$ measured by \citet{zsr13}, for a lower planet mass of $1M_{\rm{th}}$ ($\approx 0.15M_{\rm J}$ at 3 AU in the MMSN), modeled with half the numerical resolution.

\begin{figure}
\figurenum{3}
\epsscale{1.2}
\plotone{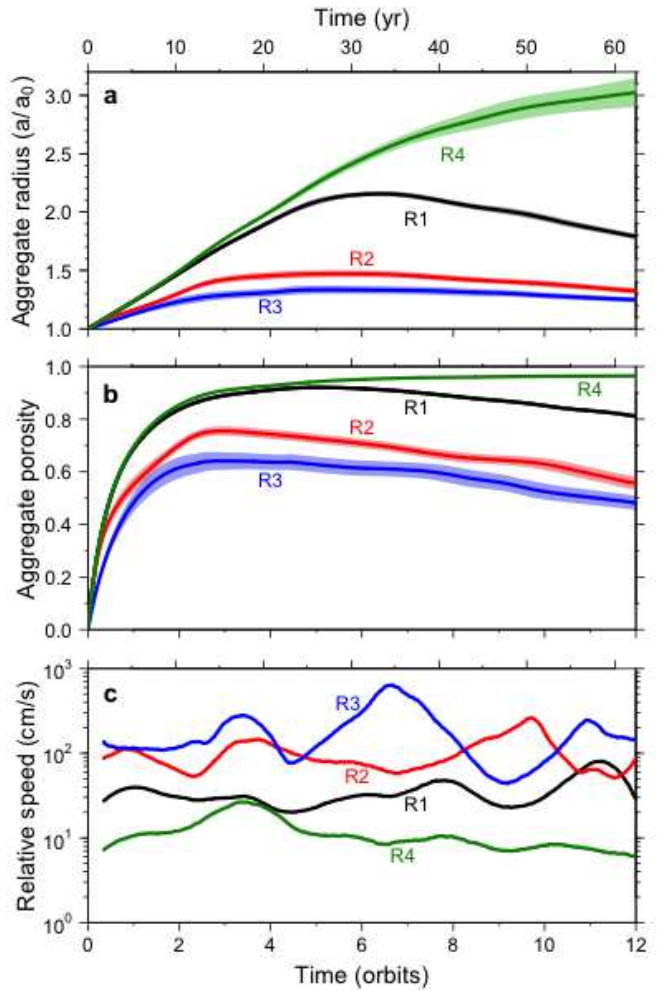}
\caption{Time evolution of \textbf{(a)} mean aggregate radius $\langle a\rangle$, normalized by the initial monomer radius $a_0$, \textbf{(b)} mass-weighted average aggregate porosity, and \textbf{(c)} simple moving average (SMA) of relative speed between the two particles involved in each collision, for the monodisperse case. The SMA is taken every 0.35 orbits. Each curve color represents the region R$n$ ($n$=1,...,4) of Fig. \ref{fig:xymap} in which coagulation was calculated. In panels \textbf{(a)} and \textbf{(b)}, the bands surrounding the solid lines denote one standard deviation of the data taken over the 128 macrocells of each region. } 
\label{fig:aggradmon}
\end{figure}

The coagulation evolution of the particle populations with an initial monomer radius of 1 $\mu$m is shown in Fig. \ref{fig:aggradmon}. Each curve color refers to one of the four regions of Fig. \ref{fig:xymap}. Figure \ref{fig:aggradmon}a shows the mean radius of the aggregates, normalized by the initial monomer radius $a_0=1$ $\mu$m. Sticking without restructuring occurs more effectively than compaction for grains subject to the flow conditions in region R4 (green curve), where gas velocities are relatively low, as can be seen in Fig. \ref{fig:radprofs}b. On the other hand, growth is stalled for particles in regions R1, R2 and R3 (black, red and blue curves, respectively), which have higher turbulent gas velocities. 

The evolution of porosity, defined by the mass-weighted average enlargement factor $\langle \psi_k\rangle_m$ as $1-1/\langle \psi_k\rangle_m$, is shown in Fig. \ref{fig:aggradmon}b. The compaction of aggregates in regions R2 and R3 is evident, and best understood by looking at the relative impact speeds of the aggregate pairs involved in each collision, as shown in Fig. \ref{fig:aggradmon}c. The highest relative speeds are generated in region R3, as expected from the magnitude of the gas velocity there (Fig. \ref{fig:radprofs}b).  
 
\begin{figure}
\figurenum{4}
\epsscale{1.2}
\plotone{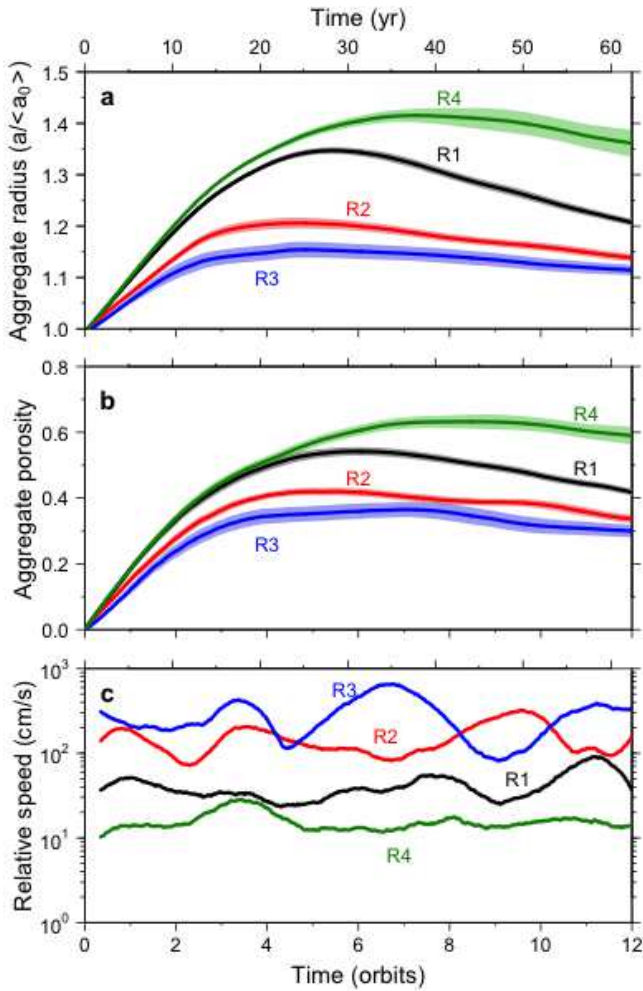}
\caption{As in Fig. \ref{fig:aggradmon}, but for the MRN-type initial dust size distribution. In panel \textbf{(a)}, the mean aggregate radius is normalized by the mean initial monomer radius. } 
\label{fig:aggradmrn}
\end{figure}

The case of an initial MRN-type size distribution is qualitatively similar to the monodisperse case, but with less overall growth. Figure \ref{fig:aggradmrn}a, which also displays the mean particle radius, shows that aggregates in all four regions have their growth reversed during the span of the calculation, with the largest growth achieved only by particles in region R4; they grow by $\sim 35$\% from the initial mean radius of 0.83 $\mu$m. The resulting dusty structures are also more compact than in the monodisperse case, with final porosities in region R3 of 30\%, compared to 48\% in the same region when $a_0=1\,\mu$m. 

\section{Discussion}\label{sec:discussion}
Our results indicate that growth by sticking occurs most effectively outside the gap (region R4), where the gas density is higher and the gas turbulent velocity is lower (Figs. \ref{fig:radprofs}a,b). In this region, we also found that the mean vertical component of the magnetic field, $B_z$, is relatively low (Fig. \ref{fig:radprofs}d). 

Region R2, located in one of the two planetary wakes, is characterized by lower densities than R4, but higher gas velocities. It is interesting that R2 and R3 (the latter one located in the middle of the low-density gap) are two regions where $B_z$ has a local maximum and a local minimum, respectively, yet the level of aggregate compaction is similar. 

The evolution of the aggregate radius exhibits different trends in the high-gas-density region R4, for the two dust populations with different initial size distributions. In the case of the initially monodisperse size distribution, the mean aggregate radius in R4 increases with time (although somewhat more slowly in the last third of the coagulation calculation), whereas in the case of the MRN-type initial distribution the mean radius decreases in the last four orbits. It is possible, then, that at least in some regions in the vicinity of the planet, different initial dust size distributions could lead to significant differences in aggregate growth rates. This point needs to be addressed in future investigations, using, for example, semi-analytical distributions obtained for coagulation/fragmentation equilibrium \citep{birn11}.

Perhaps the most salient omissions in our MHD calculation are non-ideal effects caused by the varying ionization fraction of the disk. Magnetic field lines can be easily drawn into the planet-induced gap, and the Hall effect dominates the onset of the MRI depending on the relative orientation between the vertical component of the field and the rotation axis of the disk \citep{kw15}. Nevertheless, certain combinations of column density, magnetic field strength, and dust content can render the gap susceptible to the MRI in the ideal-MHD regime, such as we assume here. 


Our use of an orbital radius of 3 AU for a Jupiter-mass planet is consistent with the Grand Tack hypothesis. Our MHD gap model can be assumed to represent a snapshot of the Grand Tack scenario. In this setting, it is instructive to look at our calculations of aggregate porosities within the context of the formation of meteorite components. In the CV chondrite Allende, chondrule rims, composed of sub-micron grains, may have accreted with high porosities of 70--80\% \citep{bland11}. If Jupiter's migration did indeed occur and was contemporaneous with the formation of meteorite parent bodies, such high porosities may have occurred outside the gap, where turbulent gas velocities were lower. In any case, further exploration of our simulation parameters is needed to determine the effect of varying magnetic field strengths and geometries, a different equation of state, and electrical charging of dust grains. The latter effect could delay the growth of rim-forming grains due to electrostatic repulsion \citep{otts11}.

\section{Conclusions}\label{sec:conclusions}

Our Monte Carlo calculations of dust growth in and around a gap opened by a Jupiter-mass planet, in the presence of turbulence generated by the magnetorotational instability, indicate that aggregate compaction is effective inside the gap, near the gap edge, and in the planetary wake. This is due to the high turbulent relative speeds between aggregates (in the range $\sim$ 20 -- 600  cm/s). In these regions, the relative kinetic energies frequently exceed the minimum energy required for restructuring. The lowest porosities occur inside the gap, with a value of $\sim 48\%$ if the initial size distribution of the population is monodisperse (with the radius of all initial monomers equal to 1 $\mu$m), and $\sim 30\%$ if the initial size distribution follows a power law with exponent -3.5, as in the MRN distribution for interstellar dust. 

The most porous aggregates occur in a high-gas-density region outside the gap, where the turbulent relative speeds are lower. In that region, porosities reach extremely high values of $\sim98\%$ for the initially monodisperse population. In the case of the MRN-type population, the porosity reaches  $\sim 63\%$.  

The outcome of our MHD model of a planet-induced gap, with the planet's mass equal to the mass of Jupiter, differs noticeably from some of the results obtained by \citet{zsr13}. In particular, the radial profile of the vertical component of the magnetic field is doubly peaked in our simulation, with one peak at either side of the planet's radial position $x/H=0$, whereas in the \citet{zsr13} study the profile exhibits only one peak at $x/H=0$, for a planet 6.7 times less massive. Future investigations need to determine the effect of the planetary mass on this radial profile, as the vertical component of the magnetic field plays a key role in the susceptibility of a gap to the MRI \citep{kw15}.

We will address the effect of grain charging in subsequent work, using a more sophisticated numerical scheme to treat the growth of dust aggregates.

\end{document}